# Two Kerr black holes with axisymmetric spins: An improved Newtonian model for the head-on collision and gravitational radiation

M.E. Araújo
*Departamento de Matemática*
*Universidade de Brasília,*
*70.910-900 Brasília - D.F., Brazil.*

P.S. Letelier and S.R. Oliveira
*Departamento de Matemática Aplicada*
*Universidade Estadual de Campinas,*
*13081-970 Campinas, S.P., Brazil.*
(September, 1997)

We present a semi-analytical approach to the interaction of two (originally) Kerr black holes through a head-on collision process. An expression for the rate of emission of gravitational radiation is derived from an exact solution to the Einstein's field equations. The total amount of gravitational radiation emitted in the process is calculated and compared to current numerical investigations. We find that the spin-spin interaction increases the emission of gravitational wave energy up to 0.2% of the total rest mass. We discuss also the possibility of spin-exchange between the holes.

## I. INTRODUCTION

The two body problem is one the most important and difficult challenges in general relativity. As a particular case, the head-on collision of two black holes has recently gained new insights from several techniques, namely, numerical [1], perturbative [2], post-Newtonian [3] and semi-analytical methods [4]. These in turn concentrated their initial efforts in studying the collision of two Schwarzschild black holes. A semi-analytical approach developed [4] to treat the Schwarzschild case was based on an exact solution of Einstein's field equations. A particular case of this solution describes a geometry that can be interpreted as a static axisymmetric spacetime with two black holes plus a conical singularity between them. It is then possible to obtain the "force" of attraction between the holes and their "acceleration". These expressions were used to calculate the rate of emission and the total amount of gravitational wave energy released in the process. The remarkable agreement between these results and those of the numerical treatment seems to indicate that the static acceleration contribution is very important for the head-on collision problem. This conjecture is emphasized by the result that, in the context of perturbation theory, the gravitational wave emission of boosted perturbed black holes is very similar to that of the static case [5].

The previous success with the semi-analytical approach strongly motivated the present development of a similar treatment of a toy model for the head-on collision problem of two Kerr black holes. We start with an exact solution of Einstein's field equations that can be interpreted as a stationary axisymmetric spacetime with two collinear Kerr black holes and a singularity on the axis of symmetry. This solution has some free parameters that determine the nature of the singularity. This singularity has been shown to be either a spinning rod between the holes or a pair of spinning strings extending from each hole to infinity, or both in the most general case [6]. For each of these cases, one can obtain a stationary "force" of attraction and the corresponding "acceleration" from which we calculate the rate of emission and the total gravitational wave energy released in the collision when we "cut" the strut or the strings and let the holes collide. It is important to point out that in the Kerr case there is also a spin-spin interaction between the holes [8], which is always a repulsive contribution to the "attractive force".

The plan of this paper is as follows: In section II we discuss the "stationary" force between the holes with axisymmetric spins. In section III the equations of motion for the holes are obtained. In section IV we derive an expression for the rate of emission of gravitational radiation and calculate the total output of gravitational wave energy released for several initial configurations. These results are summarized in one table. Section V is dedicated to the possibility of the holes' spin variation during the collision process and section VI to our conclusions.

## II. THE TWO BODY INTERACTION FORCE

The line element of a stationary axisymmetric spacetime is given by the Lewis metric [9]

$$ds^2 = \exp(2\psi)\left(dt - w\,d\phi\right)^2 - \exp(-2\psi)\left[\exp(2\nu)\left(dr^2 + dz^2\right) + r^2 d\phi^2\right] \tag{1}$$

in cilindrical coordinates $0 \leq r < \infty$, $-\infty < z < \infty$ and $0 \leq \phi < 2\pi$. The metric functions $\psi = \psi(r,z)$, $\nu = \nu(r,z)$ and $w = w(r,z)$ satisfy the vacuum Einstein's field equations:

$$\psi_{rr} + \psi_r/r + \psi_{zz} = -\frac{\exp(4\psi)}{2r^2}\left(w_r^2 + w_z^2\right), \tag{2}$$

$$w_{rr} - w_r/r + w_{zz} = -\frac{4}{r}\left(w_r\psi_r + w_z\psi_z\right), \tag{3}$$

$$\nu_r = r\left(\psi_r^2 - \psi_z^2\right) - \frac{\exp(4\psi)}{2r}\left(w_r^2 + w_z^2\right), \tag{4}$$

$$\nu_z = 2r\psi_r\psi_z - \frac{\exp(4\psi)}{2r}w_r w_z. \tag{5}$$

The metric function $\nu$ is obtained, up to a constant, by quadrature once the solutions $\psi$ and $w$ for the non-linear coupled system of equations (2) and (3) are known. These equations are also the integrability conditions for the equations (4) and (5) and they can also be obtained from the action $\int \left[\left(\psi_r^2 + \psi_z^2\right) - \frac{\exp(4\psi)}{4r^2}\left(w_r^2 + w_z^2\right)\right] r dr d\phi dz$ [10]. Asymptotically $\psi$ plays the role of a Newtonian potential and $w/4$ of a body's angular momentum per radial distance. $w$ is related to the usual twist potential $\chi$ by $r\chi_r = w_z$ and $r\chi_z = -w_r$.

Solutions of the system of equations (2)–(3) have been obtained using Bäcklund transformation and inverse scattering techniques [11]. In general these solutions have singularities. We point out that the choice of the integration constants in the solution is a crucial step towards the interpretation of the nature of the singularity and consequently of the resulting spacetime geometry. In this paper we are going to deal with two possible configurations. One that corresponds to two collinear Kerr black holes with a twisted spinning strut between them and the other case, that of a couple of spinning strings extending from each of the holes to infinity.

The strut and string's energy momentum vanishes everywhere except at the axis. Furthermore, their effective mass is null and their angular momentum per unit length is given by $w\exp(\psi)/4$ at $r = 0$ [12] [6]. Let $m_1$, $m_2$ be the masses and $a_1$, $a_2$ be the angular momenta per unit mass of the holes (hereafter called simply spins), respectively. It has shown that the compression force on the strut is given by [6]

$$F_{\text{strut}} = -\frac{m_1 m_2}{d^2 - (m_1 + m_2)^2 + (a_1 - a_2)^2}, \tag{6}$$

where $d$ is the coordinate distance between the center of the holes. We assume that this is also the "stationary force of attraction" between the holes if the strut is removed. In order to get rid off the semi-infinity strings the spacetime has the constraint $\frac{a_1}{a_2} = \frac{m_1}{m_2}$ so it is applicable to the parallel spin case only.

The anti-parallel spin case can be treated using the tension force on the string as the "stationary force of attraction" between the holes if both strings are cut. The force of attraction is given by

$$F_{\text{strings}} = -\frac{m_1 m_2}{d^2 - (m_1 - m_2)^2 + (a_1 - a_2)^2} \tag{7}$$

which is opposit to the tension on the strings.

Both forces are singular at a finite separation $d$, nevertheless our computation stops before that. Recall that $\sigma_k \equiv \sqrt{m_k^2 - a_k^2}$, $m_k \geq a_k$, $k = 1, 2$ is the coordinate distance between the center of each isolated hole and its infinite redshift surface along the axis. In this paper we consider $d^2 > (m_1 + m_2)^2 - (a_1 - a_2)^2 > (\sigma_1 + \sigma_2)^2$. Recall also that these coordinates distances correspond asymptotically to the distances measured by an observer at infinity.

Note that the spin-spin interaction reduces the attraction intensity unless the holes are co-rotating ($a_1 = a_2$). If the holes have anti-parallel spins ($a_1 = -a_2$), the spin-spin repulsion has a maximum [6]. But it gives always a repulsive contribution to the force. For large distances the spin-spin repulsion is $m_1 m_2 (a_1 - a_2)^2/d^4$.

### III. EQUATIONS OF MOTION

We now investigate the time evolution of a two-body system interacting through the force (6) applicable only to the parallel spins black holes. We assume that Newton's equations of motion are adequate to describe the black holes motion and that their spins and the masses are constants. We then have a Newtonian two-body problem that can be transformed into a central force problem with a fixed center of mass. Let $M = m_1 + m_2$ and $\mu = m_1 m_2/M$ be the total and reduced mass of the system, respectively. Let $a_R = |a_1 - a_2|$ be the relative spin of the holes. Their relative acceleration is given by

$$\frac{d^2z}{dt^2} = -\frac{M}{z^2 - M^2 + a_R^2} \ . \tag{8}$$

Let $\overline{z} = z/M$ , $a = a_R/M$ and $t \to t/M$. Recall that the spin of a hole is constrained to be less than its mass, so $a \leq 1$.

The first integral of equation (8) is

$$\tfrac{1}{2} \dot{\overline{z}}^2 + V = \varepsilon \ , \tag{9}$$

where $\varepsilon$ is the total energy per unit of reduced mass, $\dot{\overline{z}} = d\overline{z}/dt$ and the potential $V$ is

$$V(\overline{z}, a) = -\frac{1}{2\sigma} \ln \left[ \frac{\overline{z} + \sigma}{\overline{z} - \sigma} \right] \ , \tag{10}$$

where $\sigma = \sqrt{1 - a^2}$ .

We now investigate the time evolution of a two-body system interacting through the force given in (7) applicable to the anti-parallel spins black holes. We remark that this space-time is not asymptotically flat because of the semi-infinity strings. Even so we assume that Newton's equations of motion are adequate to describe the black holes motion and that the spins and the masses of the holes are constants.

Their relative acceleration is given by

$$\frac{d^2z}{dt^2} = -\frac{m_1 + m_2}{z^2 - (m_1 - m_2)^2 + a_R^2} \ . \tag{11}$$

Let $\overline{z} = z/M$ , $a = a_R/M$ , $\delta \equiv (m_1 - m_2)/M$ and $t \to t/M$. The first integral of equation (11) is similar to (9) with the potential

$$V(\overline{z}, a, \delta) = \begin{cases} -\frac{1}{2\beta} \ln \left[ \frac{\overline{z}+\beta}{\overline{z}-\beta} \right] & \text{for } \delta^2 > a^2 \\ -\frac{1}{\overline{z}} & \text{for } \delta^2 = a^2 \\ -\frac{1}{|\beta|} \left( \frac{\pi}{2} - \arctan\left( \frac{\overline{z}}{|\beta|} \right) \right) & \text{for } \delta^2 < a^2 \end{cases} \tag{12}$$

where $\beta = \sqrt{\delta^2 - a^2}$ .

Therefore we can treat both, parallel and antiparallel spins cases, using the potential (12) since $\sigma = \beta$ if we set $\delta = 1$ . If $\delta^2 = a^2$ the potential becomes the Newtonian.

### IV. EMISSION OF GRAVITATIONAL RADIATION

For the computation of the gravitational radiation luminosity (rate of emission of gravitational wave energy) we use the standard quadrupole formula [7]. In our model this is a good first approximation to get an estimate of the total amount of gravitational radiation during the head-on collision. The formula takes an average over several wave-lenghts and we integrate the luminosity along the whole collision path up to a point where the holes coalesce.

For a two particle system separated by a distance $\overline{z}$ the luminosity is given by

$$\frac{dE_g}{dt} = \tfrac{6}{45} \mu^2 \left( \dddot{\overline{z}^2} \right)^2 \ . \tag{13}$$

Our main concern is the energy emitted along the collision process. Therefore we need the rate of emission of gravitational wave energy per unit length. For a head-on collision of two bodies with equation of motion (9) the quadrupole formula (13) yields

$$\frac{dE_g}{d\overline{z}} = \frac{dE_g}{dt} \frac{1}{\dot{\overline{z}}} = \begin{cases} \frac{8}{15}\mu^2 \frac{(\overline{z}^2 - 3\beta^2)^2}{(\overline{z}^2 - \beta^2)^4} \sqrt{2\varepsilon + \frac{1}{\beta} \ln \left[ \frac{\overline{z}+\beta}{\overline{z}-\beta} \right]} & \text{for } \delta^2 > a^2 \ , \\ \frac{8}{15}\mu^2 \frac{1}{\overline{z}^4} \sqrt{2\varepsilon + \frac{2}{\overline{z}}} & \text{for } \delta^2 = a^2 \ , \\ \frac{8}{15}\mu^2 \frac{(\overline{z}^2 + 3|\beta|^2)^2}{(\overline{z}^2 + |\beta|^2)^4} \sqrt{2\varepsilon + \frac{2}{|\beta|} \left( \frac{\pi}{2} - \arctan\left( \frac{\overline{z}}{|\beta|} \right) \right)} & \text{for } \delta^2 < a^2 \ . \end{cases} \tag{14}$$

Note that, for $\delta^2 > a^2$, as $\overline{z}$ is decreasing in the collision process, $\frac{dE_g}{d\overline{z}}$ reaches a maximum at $\overline{z} \approx \sqrt{5}\beta$, then it decreases to zero at $\overline{z} = \sqrt{3}\beta$ and finally it diverges at $\overline{z} = \beta$. The behavior of the rate of emission of gravitational radiation as a function of the distance between the holes is similar to the Schwarzschild case [4]. The main difference is the fact that the two Kerr black holes horizons touch each other at $\overline{z} = (\sigma_1 + \sigma_2)/M < \beta < \sigma$ and at this point $\frac{dE_g}{d\overline{z}}$ is finite, whereas it diverges at the horizon's touching point in the Schwarzschild case. In any case, as before, the description of the head-on collision problem by this approach is valid only for $\overline{z} \geq \overline{z}_{\text{final}} \equiv \beta/\tanh\left(\frac{\beta}{2}\right)$, the point where the holes would reach the speed of light if released at the rest from infinity.

For the $\delta^2 < a^2$ case, as $\overline{z}$ is decreasing $\frac{dE_g}{d\overline{z}}$ reaches a maximum at $\overline{z} = 0$ and it is finite everywhere. Again the description is valid only for $\overline{z} \geq \overline{z}_{\text{final}} \equiv |\beta|/\tan\left(\frac{|\beta|}{2}\right)$ due to the speed limit.

The total amount of gravitational wave energy emitted is then given by

$$\Delta E_g(\overline{z}_{\text{initial}}, \overline{z}_{\text{final}}) = -\int_{\overline{z}_{\text{initial}}}^{\overline{z}_{\text{final}}} \frac{dE_g}{d\overline{z}} d\overline{z} \;,$$

where $\overline{z}_{\text{initial}}$ is the point of release from rest. We make a restriction to the cases $-1/2 < \varepsilon \leq 0$, which corresponds to $\overline{z}_{\text{final}} \leq \overline{z}_{\text{initial}} < \infty$. Our results are summarized in the table below, where the total amount of gravitational wave energy is given as a function of $\varepsilon$ and $\beta$ in units of $10^{-3} \times \mu^2/M$.

The efficiency of the emission of gravitational waves decreases with $\beta$ for $\delta^2 > a^2$, increases with $|\beta|$ for $\delta^2 < a^2$ and decreases with $\varepsilon$. Note that the anti-parallel case can emit as much as 3% of the rest energy $\mu^2/M$. We see for example that the collision of equal masses extremal ($m = a$) black holes with anti-parallel spins ($\beta = i$) can release at least 3 times more gravitational wave energy than the similar collision with parallel spins ($\beta = 1$).

## V. TIME VARIATION OF SPINS

Let us now turn to the possibility of time variation of the holes' spins. Since both strut or strings attached to the holes have angular momentum per unit length, they may affect the angular momentum of the holes along the collision path. It would be a mutual Lense-Thiring effect or the dragging of each other's inertial frame. The gravitational radiation does not carry out angular momentum in the quadrupole approximation due to the axisymetry [3]. Therefore it is reasonable to expect angular momentum conservation in the configuration.

In the parallel case the strut's total angular momentum is given by [6]

$$J_{\text{strut}} = -\frac{2m_2\,a_1\,(d + m_2 + m_1)(d - \sigma_1 - \sigma_2)}{d^2 - (m_1 - m_2)^2 + (a_1 - a_2)^2} \;.$$

The constraint $\frac{a_1}{a_2} = \frac{m_1}{m_2}$ has been used. Let the total angular momentum be $\ell \equiv J_{\text{strut}} + m_1 a_1 + m_2 a_2$. Let us assume that $\ell$ is conserved during the collision process. Then $a_1$ and $a_2$ have to change with the separation distance $d$. Let $\alpha \equiv a_1/m_1 = a_2/m_2$, $\delta \equiv m_2 - m_1$ and set $m_1 + m_2$ to one. In order to keep $\ell$ constant, $\alpha$ must be a real root of

$$\sqrt{1 - \alpha^2}\alpha = \left[\frac{\overline{z}}{(\overline{z}+1)} \frac{(\delta^2 + 2\overline{z}\delta^2 - 1)}{(1 - \delta^2)} - \frac{\delta^2(1 + \delta^2)}{(\overline{z}+1)(1 - \delta^2)}(1 - \alpha^2)\right]\alpha - 2\ell\frac{(\overline{z}^2 - \delta^2(1 - \alpha^2))}{(\overline{z}+1)(1 - \delta^2)} \quad (15)$$

For a given initial condition, $\ell$ and $\delta$ are known, then $\alpha$ is the root of (15) that changes with $\overline{z}$ from its initial value continuously. We find that $\alpha$ slightly decreases as the holes get close to each other for most of the initial conditions. For example, considering identical holes ($\delta = 0$) and setting $\ell = 0$ we find $\alpha^2 = (2\overline{z} + 1)/(\overline{z} + 1)^2$. In Fig 1 whe show the change of $\alpha$ as a function of the separation $\overline{z}$ for $\eta = 1/3$ with initial value $\alpha = 1/3$ at $\overline{z} = 40$.

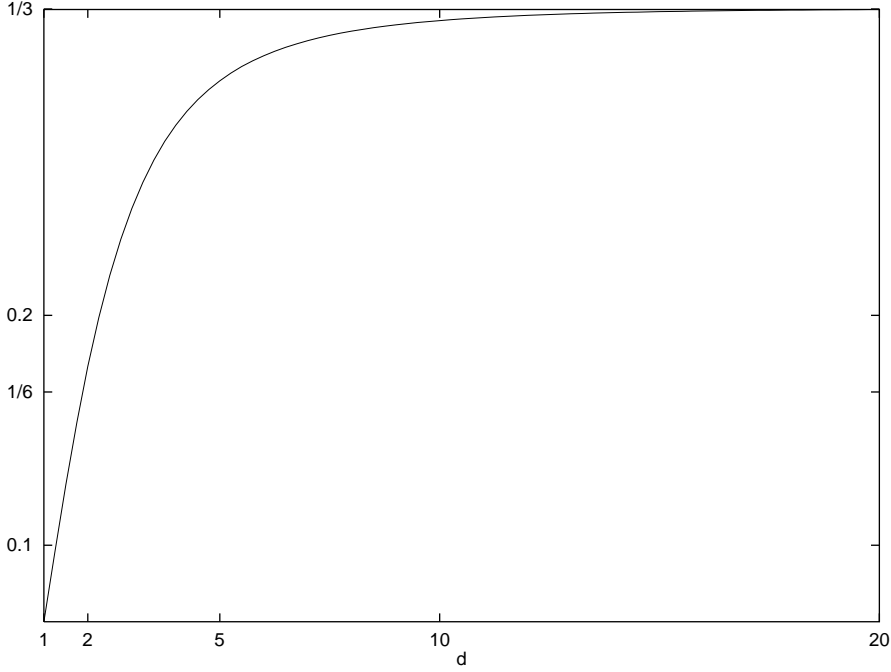

FIG. 1. Change of $\alpha$ as a function of the separation $d$ for $\eta = 1/3$ with initial value $\alpha = 1/3$ at $d = 40$.

We shall now consider the torque of the semi-infinite strings on the holes. The strings total angular momenta are not finite. Nevertheless, the rate of their change may be computed as follows

$$S_2 \, \dot{\bar{z}}_2 = m_2 \, \dot{a}_2 \tag{16}$$
$$S_1 \, \dot{\bar{z}}_1 = m_1 \, \dot{a}_1 \tag{17}$$

where $S_k$, $k = 1, 2$ are the strings spins given by [6]

$$S_1 = S_{1;2} + \frac{2 m_1 a_2 \, d}{\Delta^2}$$
$$S_2 = S_{1;2} + \frac{2 m_2 a_1 \, d}{\Delta^2}$$
$$S_{1;2} \equiv -\frac{2 m_1 m_2 a_2}{\Delta^4} \left( m_2^2 - (m_1 + d)^2 \right) + a_1 \, \left( m_1 - (m_2 + d)^2 \right) - (a_1 - a_2) \left( a_1^2 - a_2^2 \right)$$

where $\Delta^2 = d^2 - (m_1 + m_2)^2 + (a_1 - a_2)^2$. That is, we assume that each string exchanges angular momentum with its attached black hole. We use a fixed center of mass reference system in which $z_2 - z_1 = d$ and $m_1 z_1 + m_2 z_2 = 0$ and a further requirement that the holes angular momentum $\ell_H \equiv m_1 a_1 + m_2 a_2$ is kept constant along the collision path. Then the equations (16-17) are equivalent to the algebraic equations

$$m_1 S_2 - m_2 S_1 = 0 \tag{18}$$
$$m_1 a_1 + m_2 a_2 = \ell_H \tag{19}$$

Thus, $a_1(d) = a_1(d_{\text{initial}}) + A(d) - A(d_{\text{initial}})$ where $A(d)$ is the solution of (18-19), that is, it is the real root of

$$P(A) \equiv 4 \left( \delta^2 - 1 \right) A^3 - 4l \left( \delta + 2\delta^2 - 3d \right) A^2$$
$$+ \left[ 4l^2 \left( 2\delta + \delta^2 - 3d \right) + d \left( d^2 - 1 \right) (2\delta - 1) + (d+1)^2 \left( \delta^4 - \delta^2 \left( d^2 - 1 \right) - 2\delta^3 \right) \right] A$$
$$- l \left( (d+1) (\delta - 1)^2 \left( \delta^2 + d\delta - d^2 + d \right) + 4l^2 (\delta - d) \right) = 0$$

Here again we set $m_1 + m_2$ to one and $\delta \equiv m_2 - m_1$. We find that the holes spins decreases as the holes get close to each other. In Fig. 2 we show the change of spins during the collision process.

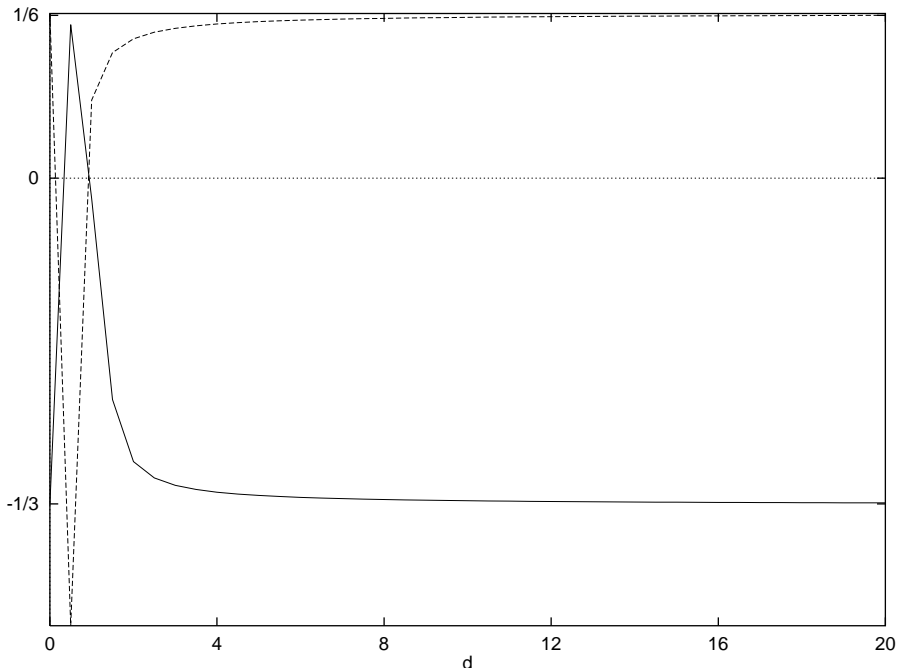

FIG. 2. Change of spins during the collision process. The settings are the same of Figure 1 except for the initial values $a_1 = -1/3$ and $a_2 = 1/6$

The change in the spins are small during most of the collision process, so we neglect the change in the gravitational radiation output presented in the Table I.

In both cases, strut and string's, we see the transfer of angular momentum from the defects to the holes. One then expects some exchange of angular momentum between the holes along the collision path.

## VI. CONCLUSIONS

We developed a semi-analytical approach of treating the head-on collision problem of two Kerr black holes. This is based on an exact solution of Einstein's field equations that can be interpreted as the configuration of two Kerr black holes with some singularities on the axis from which we obtain the attraction force between the holes. We use the force and Newtonian equations of motion to find the dynamics of the holes. We then calculate the amount of gravitational radiation released in the collision using the quadrupole radiation formula. We also discuss the possible exchange of the holes spins.

It is well known that the efficiency of conversion of rest mass into gravitational radiation in the head-on collision problem is extremely small for Schwarzschild black holes. We expect higher efficiency for Kerr black holes. If we consider two identical spinning holes far apart and compare their initial energies with the final energy of a single Kerr hole, assuming hole's angular momentum conservation, the maximum radiation efficiency is 29% for parallel spins whereas the maximum ranges between 29% and 50% for antiparallel spins [13]. Our results show that the process is more efficient for the case of Kerr black holes than for the Schwarzschild black holes, but still small. The Kerr holes emit at most 3% of the rest energy $\mu^2/M$ where $\mu$ and $M$ are the reduced and total mass respectively whereas the Schwarzschild holes emit no more than 1% [4]. Recall that $\mu/M \leq 1/4$, so that the maximum radiation efficiency we found is less than 0.2% of the total rest energy $M$.

According to the area theorems for coalescing black holes the antiparallel case can radiate more energy than the parallel case [14]. Our approach agree with the area theorems. See the first (last) columns of Table I for the gravitational wave output of parallel (antiparallel) holes in collision.

We also found that the spin-spin interaction gives a repulsive contribution to the force for both parallel and antiparallel spins. In contradistinction, according to Maxwell theory, two magnetic dipoles repel (attract) each other and, according to some authors, a spinning test particle is attracted to (repelled from) a large rotating body if they are aligned in parallel (antiparallel) [8].

We then addressed the possible exchange of spins between the holes. With simple and reasonable assumptions we conclude that the holes may spin down or spin up depending on the initial conditions and on the constant parameters of the configuration. The effect is small for most of the collision path. When the holes are close enough the spins may then change up to 100%. It is known that the spin-spin force is of post-post-Newtonian magnitude [15]. Further investigations must be done towards the clarification of this spin exchange effect.

## VII. ACKNOWLEDGMENTS

M.E. Araújo and S.R. Oliveira thank FAPDF for financial support and IMECC- Unicamp and Departamento de Matemática-UnB for their hospitality during the preparation of this work. M.E. Araújo and P.S. Letelier thank CNPq for research grants. P.S. Letelier and S.R. Oliveira also acknowledge FAPESP for financal support.

| $\varepsilon \downarrow \quad \beta \rightarrow$ | 1 | 3/4 | 1/2 | 1/4 | 0 | $i/4$ | $i/2$ | $3i/4$ | $i$ |
|---|---|---|---|---|---|---|---|---|---|
| 0 | 9.0948 | 13.1064 | 16.3233 | 18.3563 | 19.0476 | 19.7450 | 21.8679 | 25.4811 | 30.6095 |
| -0.1 | 7.4681 | 10.9436 | 13.7352 | 15.5004 | 16.1008 | 16.7065 | 18.5506 | 21.6896 | 26.1450 |
| -0.2 | 5.4081 | 8.2322 | 10.5088 | 11.9502 | 12.4408 | 12.9357 | 14.4430 | 17.0095 | 20.6521 |
| -0.3 | 3.0990 | 5.0773 | 6.6864 | 7.7086 | 8.0569 | 8.4084 | 9.4797 | 11.3052 | 13.8951 |
| -0.4 | 1.0577 | 1.9689 | 2.7258 | 3.2101 | 3.3755 | 3.5427 | 4.0526 | 4.9225 | 6.1549 |

TABLE I. In the first column are the values of $\varepsilon$. In the first line are the values of $\beta$. The other entries of the table give the total amount of gravitational radiation emmitted during the collision process.